\RequirePackage{fix-cm}
\documentclass[smallextended]{svjour3}			
\smartqed										
\usepackage{graphicx}
\begin{document}

\title{Periodic metal resonator chains for Surface Enhanced Raman Scattering (SERS)
}


\author{J. Sievers	\and
		F. Heyroth	\and 
		S. Schlenker	\and
		G. Schmidt	\and
		A. Sprafke	\and
		M. Below	\and
		C. Reinhardt	\and 
		J. Schilling$^{*}$
}


\institute{J. Sievers, M. Below, J. Schilling$^{*}$ \at Centre for Innovation Competence SiLi-Nano, Martin-Luther-University Halle-Wittenberg, Halle(Saale), Germany\\
		$^{*}$ Corresponding author; \email{joerg.schilling@physik.uni-halle.de}	
		\and
		F. Heyroth, S. Schlenker, G. Schmidt \at Interdisciplinary Center of Material Science, Martin-Luther-University Halle-Wittenberg, Halle(Saale), Germany
		\and
		A. Sprafke \at Group $\mu$MD, Institute of Physics, Martin-Luther-University Halle-Wittenberg, Halle(Saale), Germany
		\and
		C. Reinhardt \at Hochschule Bremen, Neustadtswall 30, 28199 Bremen, Germany \\
		Laser Zentrum Hannover e.V., Hollerithallee 8, 30419 Hannover, Germany}

\date{Received: date / Accepted: date}

\maketitle

\begin{abstract}
A periodic arrangement of chains of gold discs shows pronounced plasmonic grating resonances. These have a clear impact on the Surface Enhanced Raman Scattering (SERS)-signal from 4-methylbenzenethiol molecules which form self-assembled monolayers on the gold surface: Besides a clear polarisation dependence the SERS-spectra also exhibit a maximum when the excitation laser wavelength matches the plamonic grating resonance. These features are explained by a combined near and far field coupling of the individual plasmonic dipoles allowing the design of optimized nanostructures for effective SERS-substrates in the future.\par

\end{abstract}
\newpage
\section{Introduction}
\label{intro}
Although Raman spectroscopy is nowadays used as a "finger print" technique for the identification of many molecules, its big disadvantage is still the low efficiency of the inelastic Raman-Scattering process itself. However the scattering efficiency can be markedly enhanced for molecules, which are adsorbed at noble metal surfaces leading to Surface Enhanced Raman Scattering (SERS) \cite{Sers_1}. The majour reason for this is the enhanced local electric field of the light wave at specific features like tips or grooves of rough metal surfaces. However the density and shape of these features on noble metal surfaces prepared by electrochemical or other simple deposition techniques is rather arbitrary and often non-reproducible leading to a random distribution of field hot-spots with widely varying SERS-efficiency. A common goal is therefore to fabricate SERS-substrates which offer a high density of uniformly distributed hot-spots of the same intensity resulting in the same predictable SERS enhancement from all parts of the substrate area \cite{SERS-review}.\par

To specifically design substrates for increased SERS sensitivity, metal nano-structures are commonly fabricated by electron beam lithography (EBL). Das et al. for instance fabricated periodic Au nanocube structures applying EBL \cite{EBL1}. Lin et al. used focused ion beam milling (FIB) to precisely fabricate different Au disc patterns exhibiting hexagon-like, shield-like, pentagon-like, and kite-like geometries\cite{EBL2}. Furthermore Cin et al. studied the influence of the detailed geometry of the nanostructures on SERS, fabricating samples of concentric rings and arcs with EBL \cite{SERS_04}. The diversity of designs is endless when fabrication with EBL or FIB is considered and SERS enhancements up of $10^{11}$ from free standing gold bowtie-antennas were reported \cite{SERS_03}. To obtain these huge SERS enhancements the concentration of the electric dipolar near field in ultra-thin gaps between metal nanoparticle-antennas is crucial and a local plasmon resonance has to be excited. This was studied in detail for metal particle dimers \cite{nordlander-dimers}, \cite{rechberger-dimers} as well as for extended closely packed metal disc arrays \cite{bendana-arrays}. Another approach aims for the increase in Q-factor of the local plasmon resonances of the metal particles. This can be achieved by realising a strict periodic order of the metal particles resulting in sharp lattice resonances which appear at a wavelength just beyond the Wood's anomaly, when the first diffraction order becomes evanescent \cite{Barnes-1}. In this case the cancellation of radiation losses to orders higher than the zeroth reflection or transmission order results in the concentration of the light intensity within the array. This could already be used to enhance the luminescence of emitters dispersed within the metal particle array \cite{Vecchi-1}, \cite{Vecchi-2} and also an impact on SERS was already reported \cite{old-SERS-lattice}, although a large SERS-enhancement due to the lattice resonance was not yet observed \cite{Chapelle}.\par

The stated publications demonstrate clearly that the shape, size and arrangement of the nanoparticles, the material of the metal surface (like gold or silver) and the excitation wavelength strongly affect the strength of the SERS-signal. Naturally the samples fabricated by EBL or FIB only involve small areas of a few square microns since the e-beam/FIB writing process is time-consuming and expensive. However replication methods like nanoimprinting were used to transfer the patterns on larger scales paving the way to apply highly regular controlled nanostructures as general SERS substrates \cite{EBL3}.\par 

Here we report on the increase of SERS-signals from molecules attached to a one dimensional periodic array of chains of closely spaced gold discs. The idea is to achieve especially high field enhancements (intensive hot spots) by combining the effect of increased near fields in small gaps between neighbouring metal discs within the same chain with a lattice resonance generated by the periodical arrangement of the different chains.

\begin{figure}[ht]
\centering
\includegraphics[width=0.7\columnwidth]{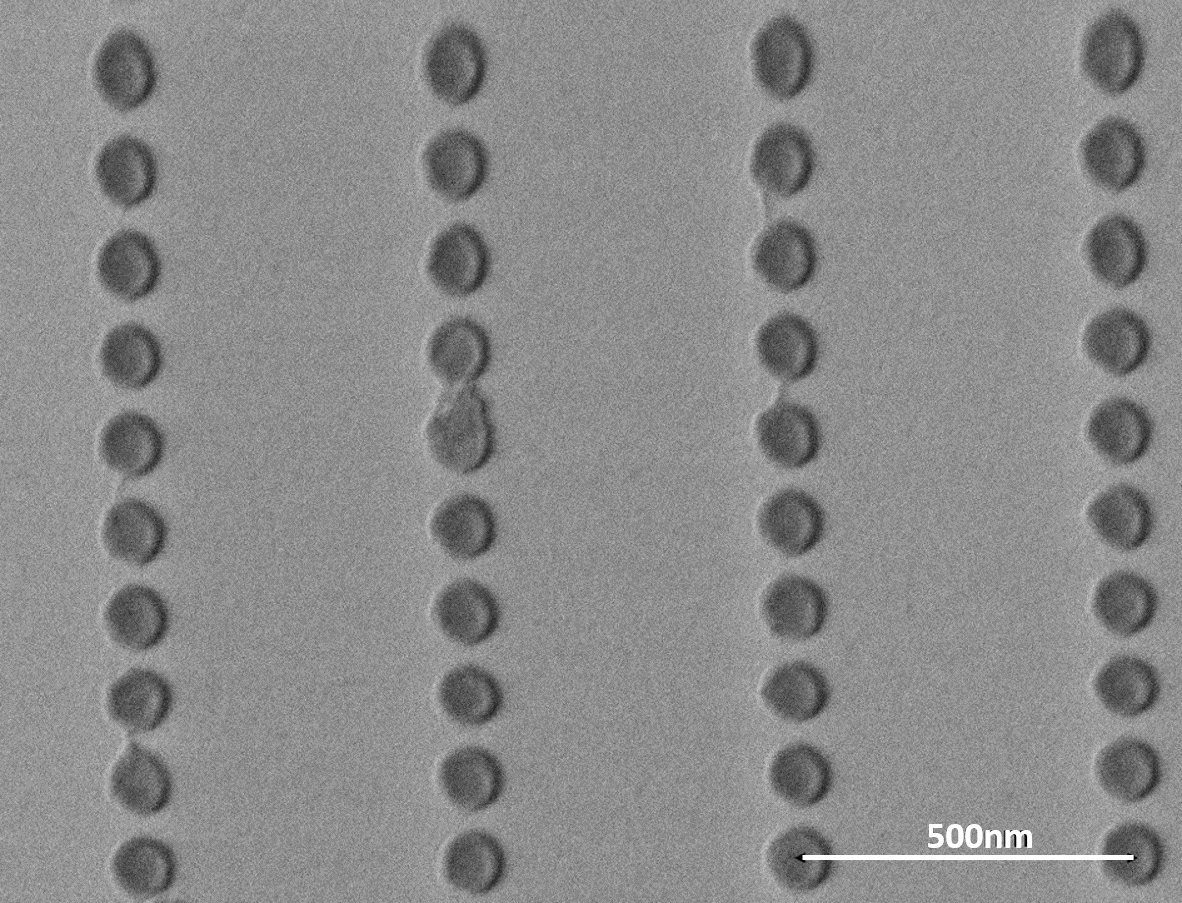}
\caption{SEM-image of an investigated structure showing the periodic arrangement of chains of gold discs on glass \label{chain_500}}
\end{figure}

\section{Plasmonic Resonances}
To realize the intended enhancement we investigate chains of gold discs, which have an individual height of 60 nm and a diameter of 65nm. The centre to centre distance between the nearest discs within the same chain is 125 nm. The period between different chains was varied between 420 to 520 nm in 20 nm increments [Fig.\ref{chain_500}]. These disc arrays with an overall dimension of $100\mu m $ x $100\mu m$ were fabricated on glass substrates applying electron beam lithography, gold evaporation (including an initially deposited 5nm Cr layer for better adhesion) and a final lift-off process.\par 

A microscope setup was used to collect the transmission spectra of the described sample employing a halogen lamp as white light source. Subsequently a collimated beam of linear polarized light was created with the help of a lens system and a polarizer. The tranmitted light was collected by an infinity corrected 20x objective and a tube lens was used to create an intermediate image of the structure in the plane of an iris. The intermediate image was then imaged onto a CCD via a plano convex lens. Adjusting the iris at the intermediate image plane and observing the image of the sample with the CCD, one could restrict the area of observation and ensure that only light transmitted through the disc chain arrays reaches the CCD. Instead of creating an image on the CCD the transmitted light could be diverted by a flip mirror and coupled into a fibre bundle, which was connected to an Acton Spectra Pro SP-2500 Monochromator with attached single channel Si-photodetector for spectral analysis.\par 

For the formation of clear plasmon grating resonances a homogeneous refractive index environment around the gold discs has to be created. We therefore applied immersion oil on the discs, which was index matched to the glass substrate resulting in a homogeneous refractive index of n=1.5 surrounding the discs.

\begin{figure}[ht]
\centering
\includegraphics[width=\columnwidth]{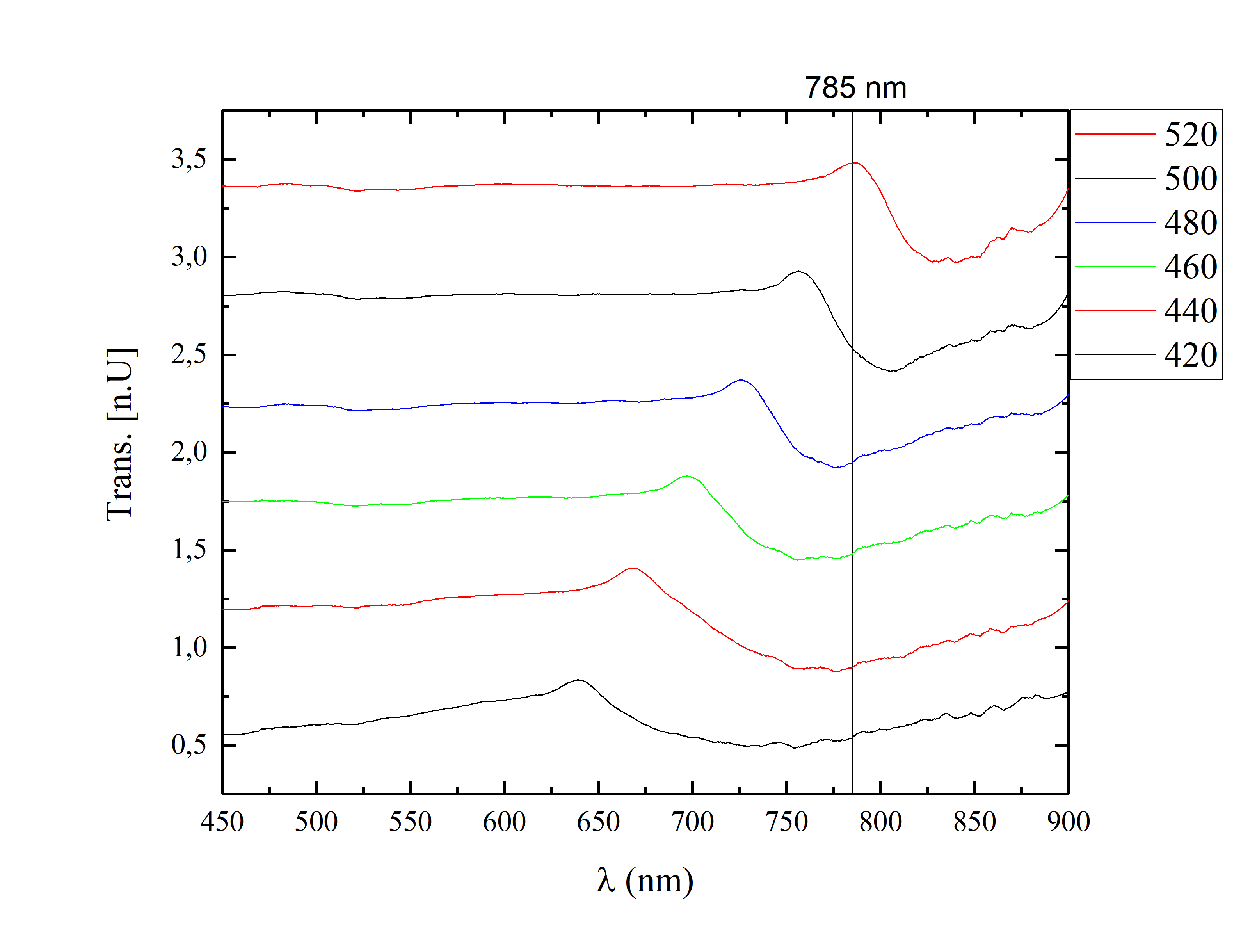}
\caption{Transmission against wavelength for different chain periods. The incident laser light is polarised parallel to the chains. The first order plasmonic grating resonance appears as a dip, which shifts to the red with increasing period between the chains. The 785nm line represents the wavelength of the Raman excitation laser used later for the SERS measurements \label{transmisson}}
\end{figure}

In Fig.\ref{transmisson} the transmission for different chain periods are plotted against the wavelength.
Every transmission spectrum shows the same characteristic shape of the curve: With increasing wavelength, we note at first a sharper peak followed by a broader dip. Increasing the period of the chains from 420nm to 520nm leads to a red shift of the dip from 740nm to 830nm [Fig.\ref{transmisson}].\par 

These dips indicate the excitation of the plasmonic grating resonances. For the chosen periods they appear as a dip on the long wavelength side of the spectrally much more extended local plasmon resonance of the single discs \cite{doktorarbeit}. The interaction between single disc resonance and the grating resonance leads to the observed Fano line shape of the described spectral peak-dip feature. In general the first order plasmonic grating resonance appears, when the incident wavelength fullfills the condition 

\begin{equation}\label{eq:1}
\lambda = n \cdot a 
\end{equation}

\noindent(n is the refractive index surrounding the discs and a is the period between the chains [nm]). In this case the scattering fields from the discs of the neighbouring chain arrive in-phase with the primary incident light wave. The constructive superposition of these electric fields leads to a strongly enhanced electron oscillation in the discs leading to an increased near field as well as enhanced absorption \cite{doktorarbeit}-\cite{schilling}.

\begin{figure}[ht]
\centering
\includegraphics[width=0.8\columnwidth]{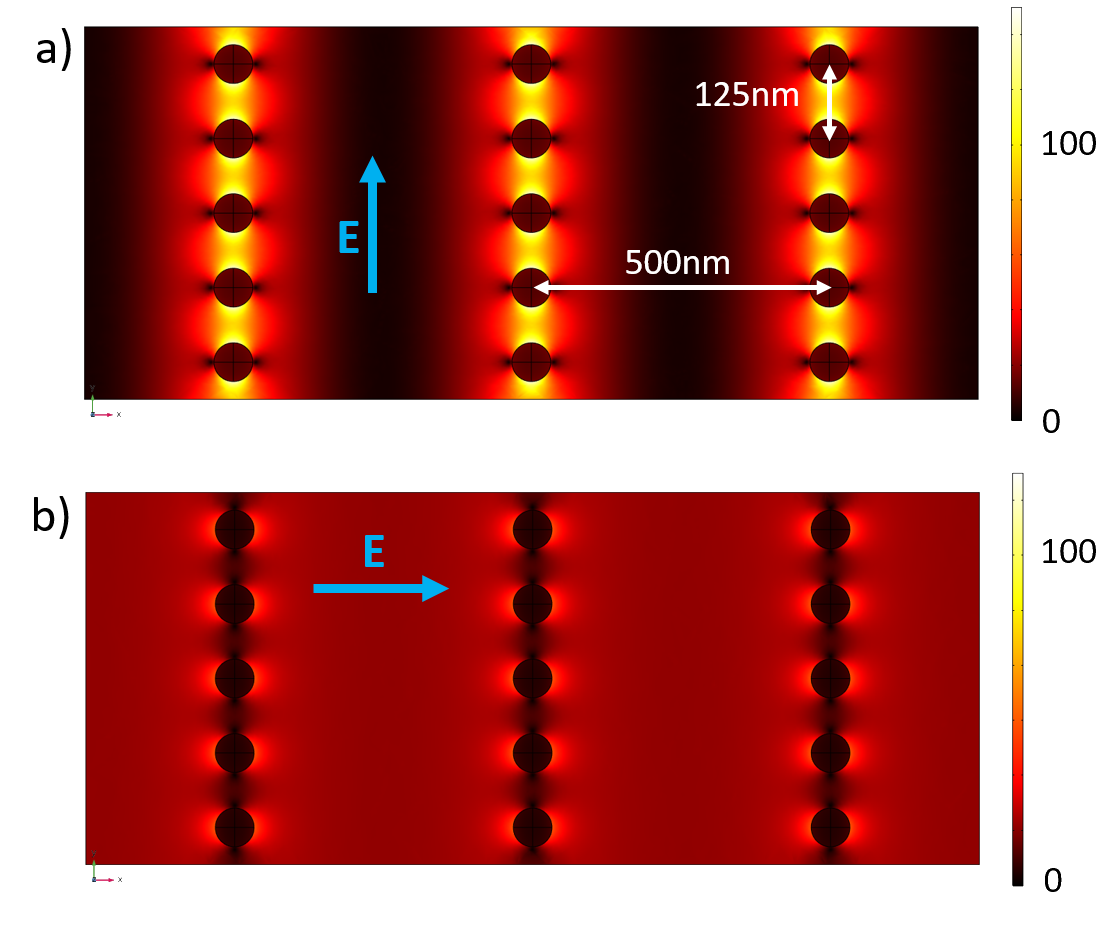}
\caption{Simulation of the local electric field around the gold discs for normal incidence at a wavelength of 785nm. a) The E-field of the incident light (blue arrow) is polarised along the chains ($0^\circ$ polarisation) and b) perpendicular to the chains ($90^\circ$ polarisation) \label{hotspot}}
\end{figure}

For SERS the enhancement of the local field at the metal surface is crucial and a finite-element simulation, applying the commercial software COMSOL, was performed to illustrate the enhancement and areas of field concentration [Fig.\ref{hotspot}]. In the first case ($0^\circ$ polarisation) the electric field and the chains are parallel to each other [Fig.\ref{hotspot}a)]. In the second case ($90^\circ$ polarisation) the electric field is perpendicularly oriented to the chains [Fig.\ref{hotspot}b)].\par 

\begin{figure}[ht]
\centering
\includegraphics[width=0.8\columnwidth]{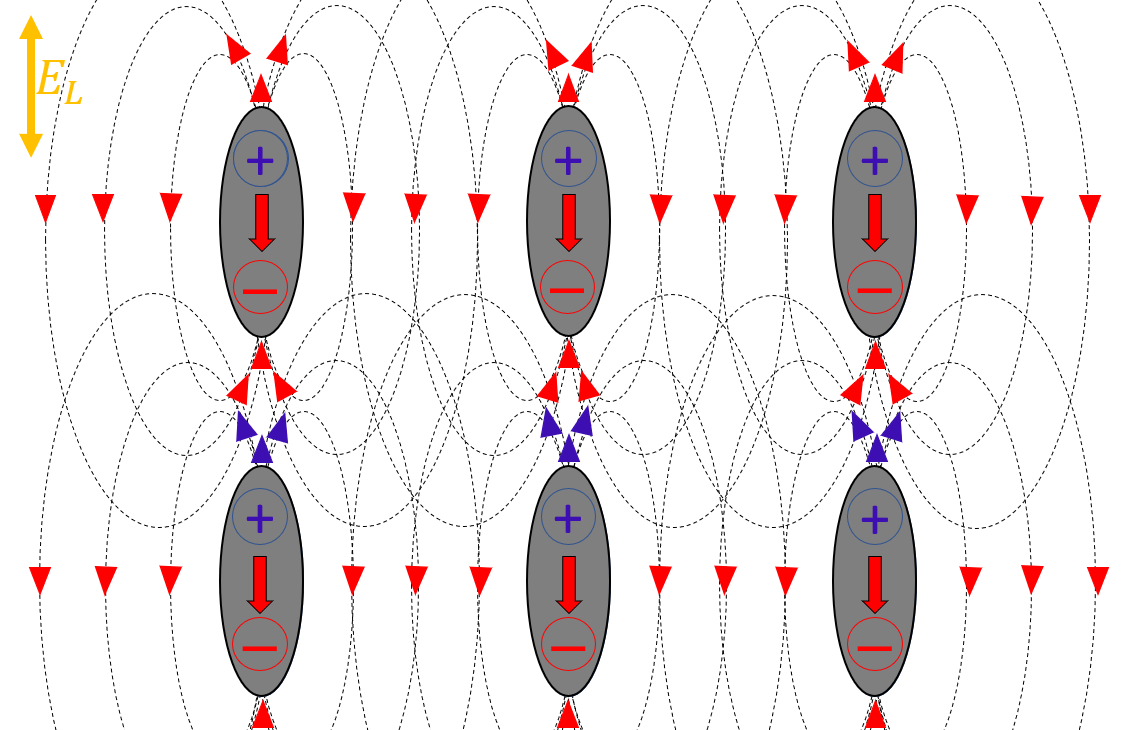}
\caption{For the case that the electric field $E_L$ of the light is polarised along the chain the dipoles form a "Head--to--Tail"--configuration. The electric field lines (dotted) of the dipoles superpose and a strong localised electric near-field is generated in the gap between the "head" and "tail" of closely spaced neighbouring dipoles resulting in field hot spots in these gaps. \label{headtotail}}
\end{figure}

The brighter colours in these heat maps indicate the areas of strong field enhancement. From this a concentration of the electric field between the neighbouring discs of the same chain is expected, when the incident light wave is polarised along the chain forming the so--called "Hot Spots" [Fig.\ref{hotspot}a)]. The increased electron oscillation at the plasmon grating resonance promises furthermore a strong enhancement of these hot spots. In contrast, for a polarisation of the incident light perpendicular to the chains [Fig.\ref{hotspot}b)] the formation of hot spots is not observed and the level of the near field appears overall subdued. 
These numerical results can already be understood considering the electric dipole fields of the discs, which are induced by the incident light. When the incident light is polarised parallel to the disc chains, the induced oscillating plasmonic dipoles of the discs form a "Head--to--Tail"--configuration [Fig.\ref{headtotail}], where the dipoles are all aligned along the chain. Since the near field of the discs is strongest at the poles of the induced dipoles, the superposition of the dipolar near fields in the head-tail configuration leads to strong electric fields in the narrow gaps between neighbouring discs of the same chain forming the already mentioned hot spots. For a polarisation perpendicular to the chains the field in the gaps between the discs of the same chain is minimal and the discs of the next chain are too far away to produce a sizeable ovelap of the near fields as the dipolar near field of each disc decays with $1/r^3$. 

\begin{figure}[ht]
\centering
\includegraphics[width=\columnwidth]{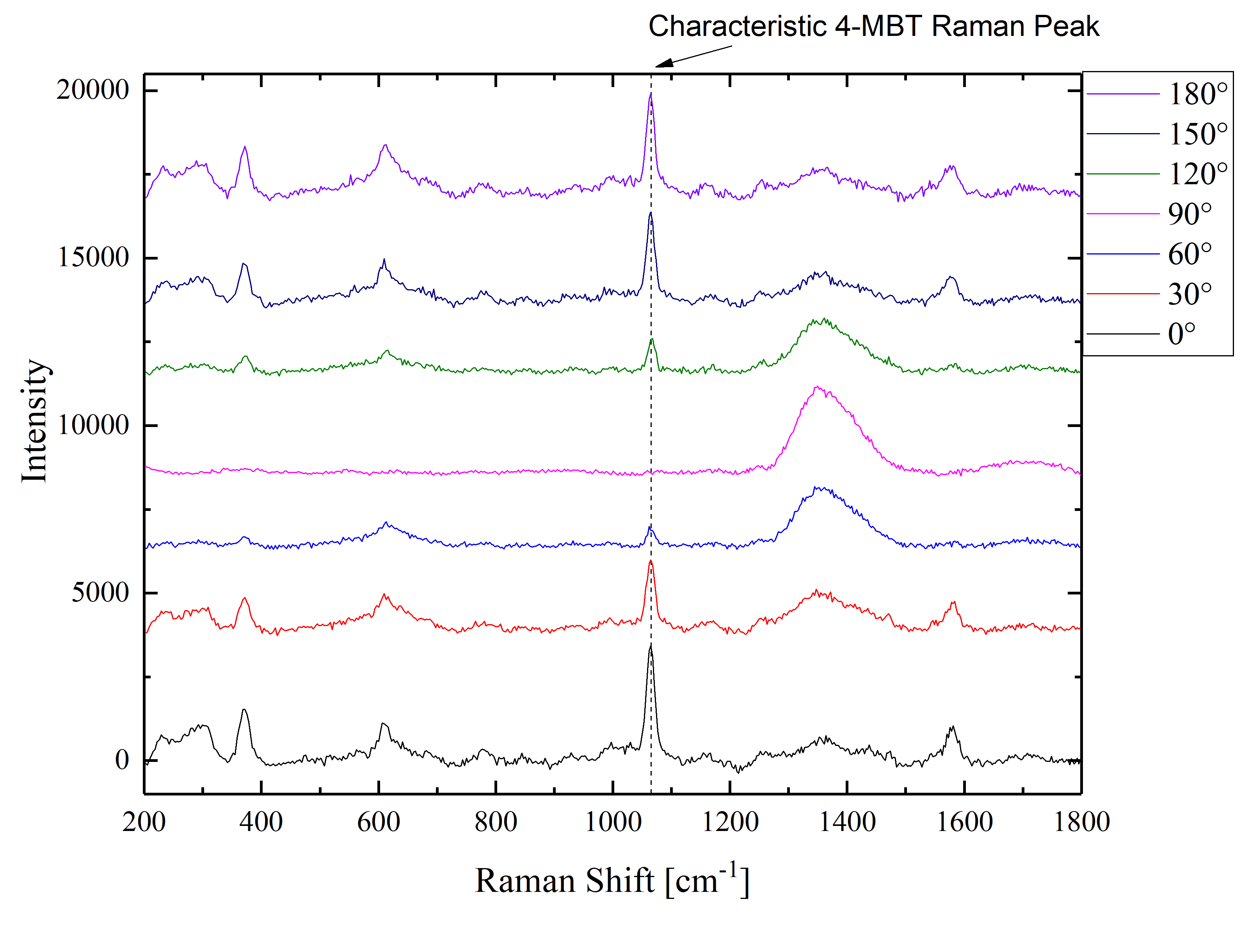}
\caption{Influence of the polarisation of the excitation laser on the SERS measurements. The intensity of the Raman-signal at $1070cm^{-1}$ varies with the polarisation of the laser. $0^\circ$ and $180^\circ$ correspond to a polarisation of the laser along the chains, $90^\circ$ for a polarisation perpendicular to the chains \label{raman_2}. Measurements performed in air.}
\end{figure}

\section{SERS results}
The predicted increased field enhancement due to the plasmonic grating resonance is of special interest for SERS measurements. The impact of this field enhancement on the Raman-Spectra of 4-methylbenzenethiol (4-MBT) was therefore investigated.\par 

To attach the 4--MBT molecules to the disc arrays, the sample was dipped into a 1mM 4--MBT/methanol solution. In the next step the excess solution was washed off using methanol and the sample was dried in air. Due to the thiol group of the 4-MBT the molecules bind directly to the gold discs and form a self assembled monolayer at the gold surface. For the SERS-measurements a Raman microscope (HORIBA Jobin Yvon LabRam HR evolution) with a laser excitation wavelength of 785nm and a power of ca. 7mW was chosen. This relatively long wavelength was selected, as gold exhibits lower losses in the near IR improving the plasmonic response of the discs, which ultimately results in higher field enhancements.\par

A 50x objective was used to focus the laser onto the sample and to collect the backscattered light and the Raman-spectrum was detected by a CCD-camera. To reduce the radiation damage to the individual molecules and measure the average enhancement of SERS in our disc array, the laser was scanned over four separate 30x30$\mu m$ squares within the $100\mu m $ x $100\mu m$ disc arrays using the DUOSCAN-mode of the Raman microscope. Finally the average of the four scans was taken. 
\par 

Fig.\ref{raman_2} illustrates the intensity of the SERS--Signal for different polarisations of the excitation laser. To keep the initial complexity of the spectra to a minimum the measurements were recorded in air. The polarisation is defined as angle between the electric field vector of the incident Raman laser and the direction of the disc chains. We concentrate on the characteristic Raman--peak at $1070cm^{-1}$, which corresponds to a wavelength of 851 nm and is caused by a combination of the C--H in plane bending and the ring breathing vibrational modes of the 4-MBT-molecule \cite{MBT-ramanlines}. The SERS--signal reaches its maximum for polarisations $0^\circ$ and $180^\circ$. A rotation of the polarisation leads to a decrease of the signal until it vanishes at a polarisation of $90^\circ$. As the measured SERS intensity depends strongly on the electric field at the surface of the discs (where the molecules are attached) the observed polarisation dependence of the SERS-signal can be attributed to the polarisation dependence of the formation of the near-field hot spots [Fig.\ref{hotspot}]. The maximum SERS signal for parallel polarisation ($0^\circ$ polarisation) is caused by the enhanced near field in the narrow gaps in head-tail dipole configuration. Vice versa the minimum SERS signal at perpendicular polarisation ($90^\circ$ polarisation) is the result of the corresponding minimal near field in the narrow gaps between the discs.\par 

In this way the observed polarisation dependent SERS-signal already implies, that the SERS signal mainly stems from the described hot spots between the discs, which only form at parallel polarisation.\par 

\begin{figure}[ht]
\centering
\includegraphics[width=\columnwidth]{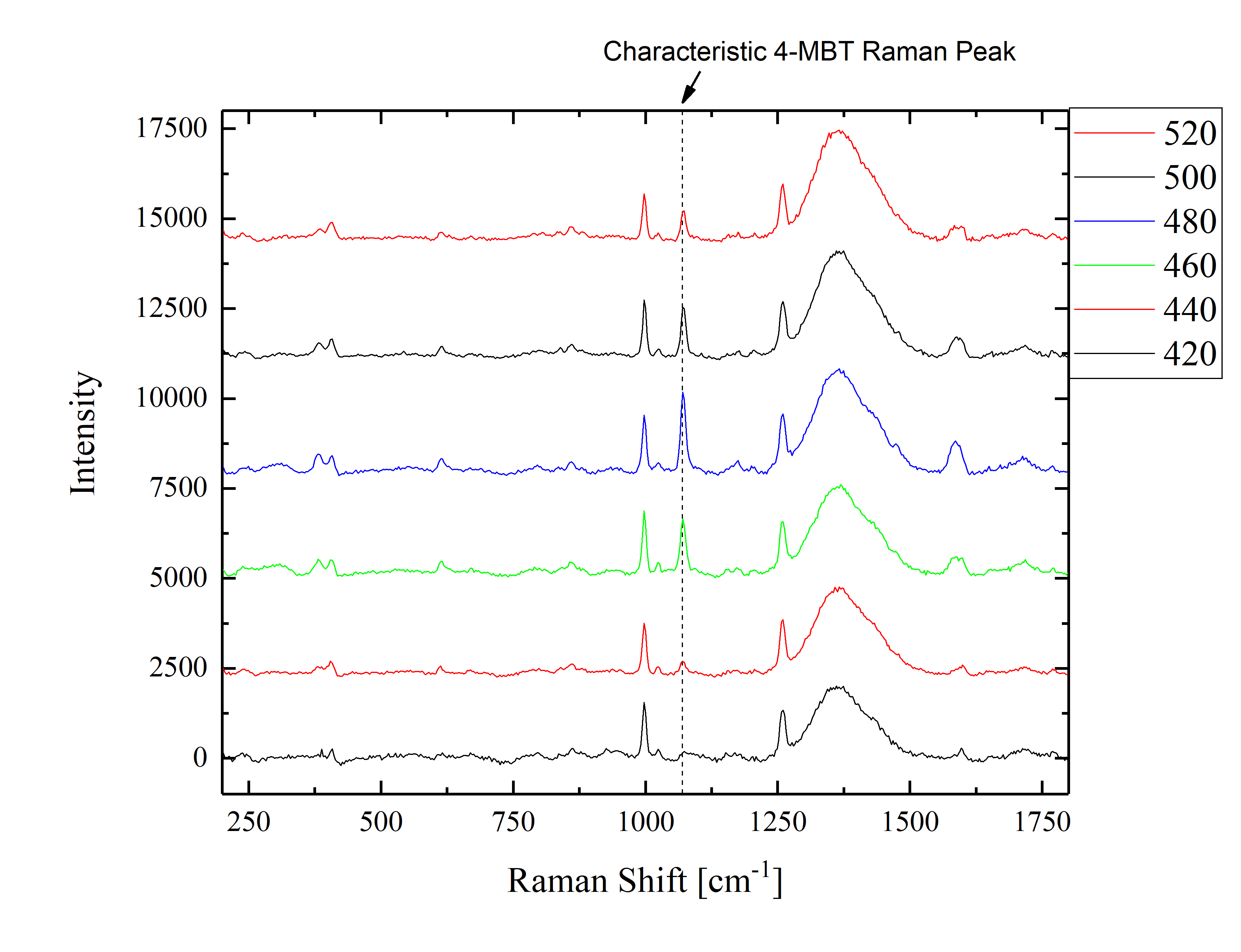}
\caption{Influence of the chain period on the SERS measurements. The intensity of the Raman-signal at $1070cm^{-1}$ varies with chain periodicity. For the measurements an immersion oil was applied on the sample with a refractive index $n=1.5$. \label{raman_1}}
\end{figure}

In a second experiment the impact of the grating resonances on the SERS-signal for different periods between the disc chains was investigated [Fig.\ref{raman_1}]. For the measurements an immersion oil with a refractive index n=1.5 was applied on the sample to create the homogeneous refractive index environment around the gold discs, which is necessary for the formation of the distinct grating resonances. Here only the parallel polarisation was used for the measurements. We concentrate again on the characteristic peak at $1070cm^{-1}$. The data show, that it reaches its maximum intensity, when the period is about 480 nm and decreases for smaller and larger periods. 
This behaviour can be ascribed to the appearance of the plasmonic grating resonances. For a chain period of 480 nm the plasmonic grating resonance matches the laser wavelength of 785nm (which is used to excite the Raman signal), so that a maximum field enhancement of the exciting laser field is achieved in the hot spots.

\section{Discussion}
\begin{figure}[ht]
\centering
\includegraphics[width=\columnwidth]{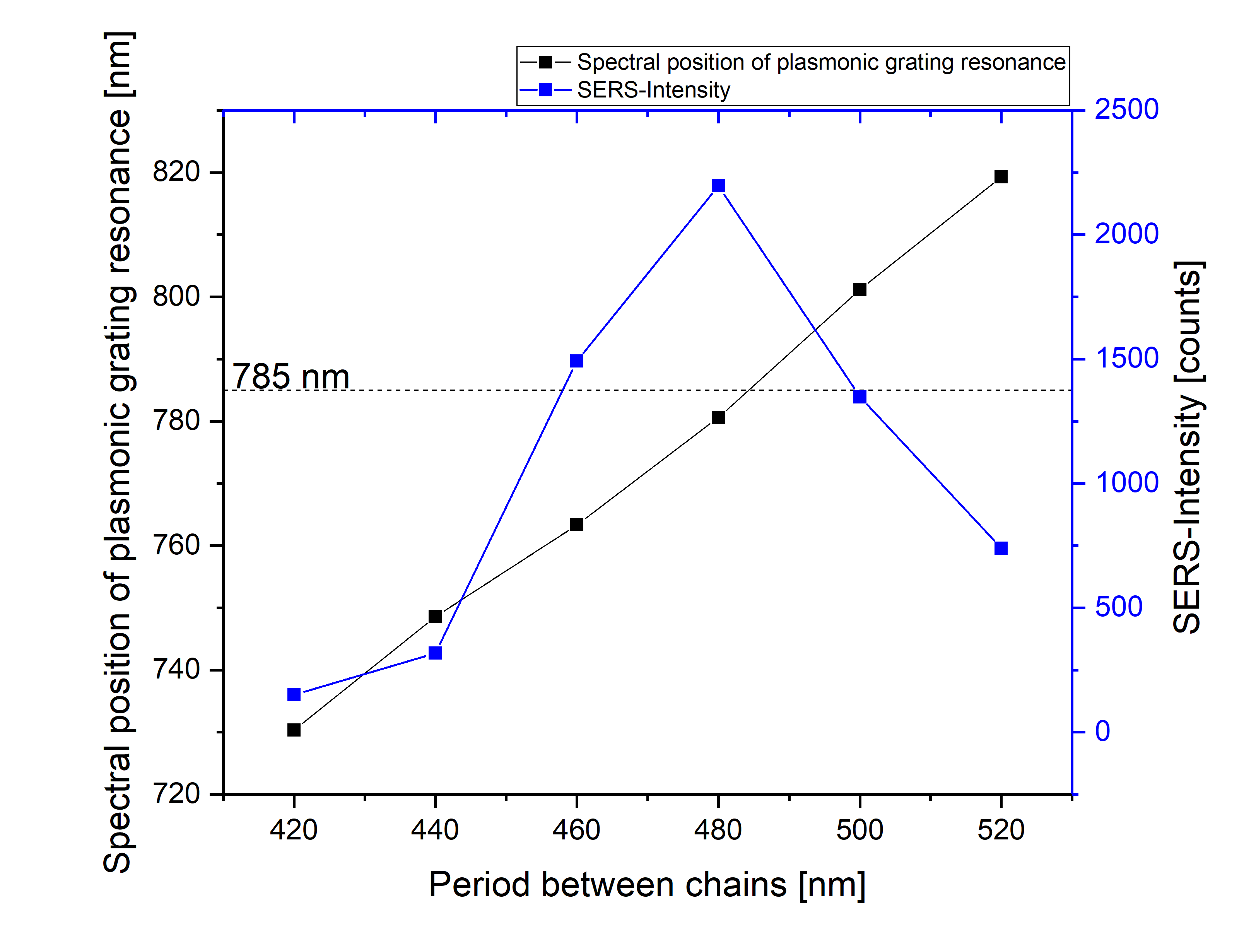}
\caption{Comparison of the dependence of the plasmonic grating resonance (black) and the SERS intensity (blue) on the chain period. The maximum SERS signal is obtained when the grating resonance occurs at the Raman laser wavelength of 785 nm for a 480nm period. \label{comparison}}
\end{figure}

The observation, that the maximum SERS-signal is obtained when the excitation wavelengths matches the grating resonance, is supported by a comparison of the spectral position of the plasmonic grating resonance and the intensity of the $1070cm^{-1}$ SERS-peak for the varying chain period [Fig.\ref{comparison}]. Here the SERS-Intensity (blue) and the spectral position of the plasmonic resonance (black) are plotted against the chain period. The plasmonic resonance for the different chain periods was read from Fig.\ref{transmisson} and shows the linear dependence on the chain period as described in (\ref{eq:1}). The wavelength of the laser is marked as dotted horizontal line. If the plasmonic grating resonance approaches the Raman excitation laser wavelength of 785nm, the SERS-Signal becomes maximum. This is the case for a period of 480 nm as pointed out before. With this, Fig.\ref{comparison} clearly demonstrates the coincidence of plasmonic grating resonance, the associated maximum field enhancement and the resulting maximum SERS-intensity. The maximum local electric field is achieved by a combination of near and far field enhancement in the disc array. For this to work, the different polarisation dependences of dipolar near and far fields is important. While for$0^\circ$ polarisation the near field is strongest parallel to the dipole oscillation along the chains , the radiating far field (which is responsible for the formation of the plasmonic grating resonances) is strongest perpendicular to the chains. With these two coupling mechanisms in mind the parameters of the disc array parameters were chosen: Narrow gaps between neighbouring discs along the chain for the enhancement of near fields, a larger period between the chains matching the wavelength requirement for the formation of the plasmonic grating resonance.\par 

A similar strategy was already employed by Crozier et al when relatively sparse arrays of optical antennas were investigated which were interlaced with a gold strip grating \cite{Crozier-1}, \cite{Crozier-2}. There the near field enhancement was realised within the narrow feed gap of the optical antennas, while the diffractive far field coupling was achieved by the strip grating. The high SERS-enhancement reported by Crozier was additionally boosted by the excitation of surface plasmon standing waves on an underlying gold layer. The potential advantage of our chain arrays compared to the sparse antenna arrays of Crozier et al. lies in the higher density of near field hot spots in our chains. The disc period within our chain is only 125nm while the spacing between the corresponding optical antennas in \cite{Crozier-1}, \cite{Crozier-2} is about 6 times larger (730nm). However, to harvest this potential, the gaps between the neighbouring discs in the same chain have to be further reduced from the current 60nm down towards 20nm or less, so that the near field enhancement in the hot spots can be further increased.

\begin{figure}[ht]
\centering
\includegraphics[width=\linewidth]{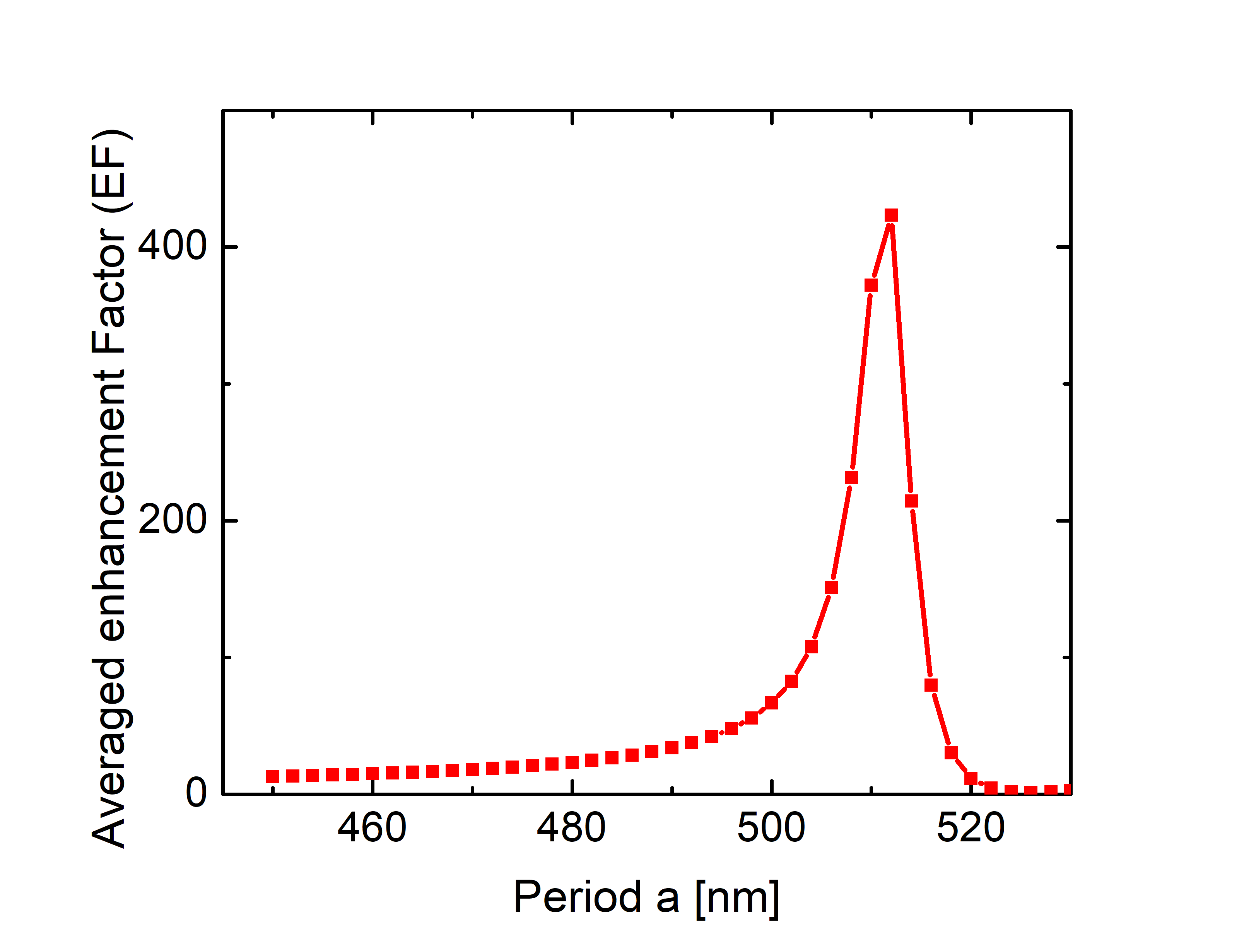}
\caption{COMSOL-Simulation: Averaged local Enhancement Factor (EF) as a function of the chain period a. The maximum EF is obtained at the chain period 510 nm. The EF decreases strongly with small variation of the chain period\label{comsol-simulation}}
\end{figure}

Moreover, based on more detailed finite element simulations we also suspect, that the maximum possible SERS intensity enhancement for the existing structure type (with the still relatively large gaps of about 60nm) might not have been reached experimentally yet. Fig. \ref{comsol-simulation} illustrates the dependence of the averaged Raman enhancement factor (EF) with respect to the chain period considering finer increments in the chain period. For this the electric field distribution within the array was simulated for a plane wave at normal incidence at the laser wavelength (785nm) and at the Raman wavelength (851nm). The EF was then calculated by means of the formula \cite{comsol}
\begin{equation}
EF = \frac{1}{A_{Ges}}\int \frac{|E_{Lloc}|^2 \cdot |E_{Rloc}|^2 }{|E_{L0}|^2 \cdot |E_{R0}|^2} \mathrm{d}A
\end{equation}
where $A_{Ges}$ is the surface area of the gold discs, $E_{Lloc}$ is the local electric field at the incident laser wavelength at the gold disc surface and $E_{Rloc}$ the local field at the raman wavelengths, while $E_{L0}$ and $E_{R0}$ are the fields of the incident light. The EF was evaluated for periods ranging from 420nm to 520nm and the simulations predict a sharp maximum of the field enhancement at a chain period of 510 nm. At this period the simulation predicts an efficient excitation of the grating resonance by the laser light.\par 

These results from finite element simulations show, that the local field enhancement is very sensitive to the exact period and a 10nm deviation has already a decisive impact on it. A finer experimental tuning of the chain period with a step size of 10 nm or 5 nm would therefore be necessary to reliably determine the ultimate maximum possible SERS enhancement in this structure. The slight difference between the theoretically predicted maximum SERS enhancement for a chain period of 510nm and the experimentally observed maximum at a chain period of 480nm is attributed to slight deviations of the experimentally realized structure from the calculated one.\par 

In conclusion we have shown, that the SERS-signal in the investigated periodic gold disc structures can be strongly increased by the correct choice of the light polarisation and the periodic distance between the chains. The results also demonstrate, that the field enhancement in our nanostructured substrates can be engineered effectively by combining near field superposition and far field interference. This might pave the way to further intentionally designed SERS substrates employing a combination of different field enhancement effects. 

\begin{acknowledgements}
The authors would like to acknowledge the funding under EFRE project ZS/2016/04/78121 and DFG-project RE3012/2.
\end{acknowledgements}

%
%


\begin{thebibliography}{}
%
%
\bibitem{Sers_1}
B. Sharma, R. R. Frontiera, A. Henry, E. Ringe, and R. P. Van Duyne, SERS: Materials, applications, and the future, Materialstoday. vol. 15, pp. 16-25 (2012)

\bibitem{SERS-review}
A. Balcytis, Y. Nishijima, S. Krishanomoorthy, A. Kuchmizhak, P.R. Stoddart, R. Petruskevicius, S. Juodkazis, From Fundamental toward Applied SERS: Shared Principles and Divergent Approaches, Adv. Optical Mater., vol.6, no. 1800292 (2018)

\bibitem{EBL1}
G. Das, M. Chirumamilla, A. Toma, A. Gopalakrishnan,
R. P. Zaccaria, A. Alabastri, M. Leoncini, and E. Di Fabrizio, Plasmon based biosensor for distinguishing different peptides mutation states, Scientic rep., vol.3, no. 1792 (2013)

\bibitem{EBL2}
Y. Lin, J. Liao, Y. Ju, C. Chang, and A. Shiau, Focused ion beam-fabricated au micro/nanostructures used as a surface enhanced raman scattering-active substrate for trace detection of molecules and inuenza virus, Nanotechnology, vol. 22, no. 185308 (2011)

\bibitem{SERS_04}
N. A. Cinel, S. Cakmakyapan, S.Butun, G. Ertas, E. Ozbay, E-Beam lithography designed substrates for surface enhanced Raman spectroscopy, Photonics and Nanostructures - Fundamentals and Applications, vol.15, pp. 109-115 (2015)

\bibitem{SERS_03}
N. A. Hatab, C.-H. Hsueh, A. L. Gaddis, S. T. Retterer, J.-H. Li, G. Eres, Z. Zhang, and B. Gu,
Free-Standing Optical Gold Bowtie Nanoantenna with Variable Gap Size for Enhanced Raman Spectroscopy, Nano Lett., vol.10, pp. 4952-4955 (2010)

\bibitem{nordlander-dimers}
 P. Nordlander, C. Oubre, E. Prodan, K. Li, M.I. Stockman, Plasmon Hybridization in Nanoparticle Dimers, Nanolett., vol.4, pp. 899-903 (2004)

\bibitem{rechberger-dimers} 
W. Rechenberger, A. Hohenau, A.Leiner, J.R. Krenn, B. Lamprecht, F.R. Aussenegg, Optical properties of two interacting gold nanoparticles, Opt. Comm, vol. 220, pp. 137-141 (2003)

\bibitem{bendana-arrays}
X.M. Bendana, G. Lozano, G. Pirruccio, J.G. Rivas, F.J.G. de Abajo, Excitation of confined modes on particle arrays, Opt. Exp., vol. 21, pp. 5636-5642 (2013) 

\bibitem{Barnes-1}
 B. Auguie, W.L. Barnes, Collective resonances in gold nanoparticle arrays, Phys. Rev. Lett., vol. 101, no. 143902 (2008)

\bibitem{Vecchi-1}
G. Vecchi, V. Giannini, J.G. Rivas, Shaping the fluorescent emission by lattice resonances in plasmonic crystals of nanoantennas, Phys. Rev. Lett., vol.102, no.146807 (2009)

\bibitem{Vecchi-2} 
G. Vecchi, V. Giannini, J.G. Rivas, Surface modes in plasmonic crystals induced by diffractive coupling of nanoantennas, Phys. Rev. B, vol. 80, no. 201401, (2009)

\bibitem{old-SERS-lattice} 
K.T. Carron, W. Fluhr, M. Meier, A. Wokaun, H.W. Lehmann, Resonances of two-dimensional particle gratings in surface-enhanced raman scattering,
Journal of the Optical Society of America, vol. 3, pp. 430-440 (1986)

\bibitem{Chapelle} 
R. Gillibert, F. Colas, R. Yasukuni, G. Picardi, M.L. de la Chapelle, Plasmonic Properties of Aluminum Nanocylinders in the Visible Range, Journal of Physical Chemistry C, vol.121, pp. 2402-2409 (2017)

\bibitem{EBL3}
S. J. Barcelo, A. Kim, W. Wu, and Z. Li, Fabrication of deterministic
nanostructure assemblies with sub-nanometer spacing using a nanoimprinting
transfer technique, ACS nano, vol.6, pp.6446-6452 (2012)

\bibitem{doktorarbeit}
N. Sardana, V. Talalaev, F. Heyroth, G. Schmidt, C. Bohley, A. Sprafke and J. Schilling, Localized surface plasmon resonance in the IR regime, Opt. Exp., vol. 24, pp.254-261 (2016) 

\bibitem{schilling}
B.B. Rajeeva, L. Lin and Y. Zheng, Design and applications of lattice plasmon resonances, Nano Research, vol.11, pp. 4423-4440 (2018)

\bibitem{MBT-ramanlines}
P.H.C. Camargo, M. Rycenga, L. Au, Y. Xia, Isolating and Probing the Hot Spot Formed between Two Silver Nanocubes, Angew. Chem. Int. Ed, vol. 48, pp. 2180-2184 (2009)

\bibitem{Crozier-1} 
D. Wang, W. Zhu, Y. Chu, K.B. Crozier, High Directivity Optical Antenna Substrates for Surface Enhanced Raman Scattering, Adv. Mat., vol. 24, pp. 4376-80 (2012)

\bibitem{Crozier-2}
W. Zhu, D. Wang, K.B. Crozier, Direct Observation of Beamed Raman Scattering, Nanolett., vol.12, pp. 6235-6243 (2012)

\bibitem{comsol}
E.C. Le Ru and P.G. Etchegoin, Rigorous justification of the $|E|^4$ enhancement factor in Surface Enhanced Raman Spectroscopy, Chemical Physics Letters, vol. 423, pp. 63-66, (2006)
\end{thebibliography}


\end{document}